\documentclass[aps, prl, twocolumn, superscriptaddress,longbibliography]{revtex4-2}
\usepackage{graphicx}
\usepackage{color}
\usepackage{textcomp}
\usepackage{bm}
\usepackage{float}
\def\bra#1{\mathinner{\langle{#1}|}}
\def\ket#1{\mathinner{|{#1}\rangle}}
\usepackage{amsmath,amssymb,amsthm}
 
\begin{document}
\title{Strong Spin-Motion Coupling in the Ultrafast Dynamics of Rydberg Atoms}
\author{V. Bharti}
\email{vineet.bharti@bristol.ac.uk}
\altaffiliation[Present address: ]{Quantum Engineering Technology Laboratories and School of Physics, Univeristy of Bristol, BS8 1TL, United Kingdom}
\affiliation{Institute for Molecular Science, National Institutes of Natural Sciences, Okazaki 444-8585, Japan}
\author{S. Sugawa} 
\affiliation{Institute for Molecular Science, National Institutes of Natural Sciences, Okazaki 444-8585, Japan}
\affiliation{SOKENDAI (The Graduate University for Advanced Studies), Okazaki 444-8585, Japan}
\affiliation{Department of Basic Science, The University of Tokyo, Meguro-ku, Tokyo, 153-8902, Japan}
\author{ M. Kunimi}
\affiliation{Department of Physics, Tokyo University of Science, Shinjuku-ku, Tokyo, 162-8601, Japan}
\author{V. S. Chauhan}
\affiliation{Institute for Molecular Science, National Institutes of Natural Sciences, Okazaki 444-8585, Japan}
\author{T. P. Mahesh} 
\affiliation{Institute for Molecular Science, National Institutes of Natural Sciences, Okazaki 444-8585, Japan}
\affiliation{SOKENDAI (The Graduate University for Advanced Studies), Okazaki 444-8585, Japan}
\author{M. Mizoguchi} 
\affiliation{Institute for Molecular Science, National Institutes of Natural Sciences, Okazaki 444-8585, Japan}
\author{T. Matsubara} 
\affiliation{Institute for Molecular Science, National Institutes of Natural Sciences, Okazaki 444-8585, Japan}
\author{T. Tomita} 
\affiliation{Institute for Molecular Science, National Institutes of Natural Sciences, Okazaki 444-8585, Japan}
\affiliation{SOKENDAI (The Graduate University for Advanced Studies), Okazaki 444-8585, Japan}
\author{S. de Léséleuc} 
\email{sylvain@ims.ac.jp}
\affiliation{Institute for Molecular Science, National Institutes of Natural Sciences, Okazaki 444-8585, Japan}
\affiliation{SOKENDAI (The Graduate University for Advanced Studies), Okazaki 444-8585, Japan}
\author{K. Ohmori} 
\email{ohmori@ims.ac.jp}
\affiliation{Institute for Molecular Science, National Institutes of Natural Sciences, Okazaki 444-8585, Japan}
\affiliation{SOKENDAI (The Graduate University for Advanced Studies), Okazaki 444-8585, Japan}

\date{\today}

\begin{abstract}
Rydberg atoms in optical lattices and tweezers is now a well established platform for simulating quantum spin systems. However, the role of the atoms' spatial wavefunction has not been examined in detail experimentally. Here, we show a strong spin-motion coupling emerging from the large variation of the interaction potential over the wavefunction spread. We observe its clear signature on the ultrafast many-body nanosecond-dynamics of atoms excited to a Rydberg $S$ state, using picosecond pulses, from an unity-filling atomic Mott-insulator. We also propose a novel approach to tune arbitrarily the strength of the spin-motion coupling relative to the motional energy scale set by trapping potentials. Our work provides a new direction for exploring the dynamics of strongly-correlated quantum systems by adding the motional degree of freedom to the Rydberg simulation toolbox.

\end{abstract}
\maketitle	

Quantum simulation platforms, such as ion crystals \cite{Monroe2021}, polar molecules \cite{Langen2023}, ultracold neutral atoms \cite{GRB17}, and Rydberg atoms \cite{BRL20}, offer remarkable opportunities to study various many-body problems, of which one important category are localized spin models, see e.g.~\cite{Ross2022,YMG13,Ketterle2020,Semeghini2021,Chen2023}. 
To mimic pure spin systems, two energy levels in the internal degrees of freedom (d.o.f.) are identified as an effective spin-1/2, and approximations are then applied onto the full Hamiltonian describing a given experimental platform, notably to decouple the external motional d.o.f. (position and momentum) from the spin dynamics. 
Recently, new proposals are emerging to purposely use \textit{spin-motion coupling} (i.e., a state-dependent force) and open new regimes of quantum simulation with Rydberg atoms~\cite{Gorshkov2019,Lesanovsky2020,Lesanovsky2020b,Lesanovsky2023,Mehaignerie2023,ZYS22}. 
In this work, based on the ultrafast Rydberg quantum platform~\cite{CTM21,BSM23}, we report on the experimental realization of an extreme regime of spin-motion coupling $\kappa$ which is (i) comparable to the spin-spin interaction strength $V$, and (ii) overly dominates the natural motional energy scale $\omega$ set by a trapping potential. We also propose a novel experimental approach, \textit{ultrafast stroboscopic Rydberg excitation}, to tune the ratio $\kappa/\omega$ over many orders of magnitude. 

Rydberg atoms display interactions ranging up to the GHz-scale at micrometer inter-atomic distances $r$~\cite{SWM10, BRL20}. The potential $V(r)$ typically follows a $1/r^3$-dependence for resonant dipole-dipole interaction, or a $1/r^6$-potential in the non-resonant van der Waals (vdW) regime.
Over the last decade, spin models have been implemented with Rydberg atoms in a gas phase~\cite{TSG16,SchleierSmith2020,Signoles2021}, in an optical lattice~\cite{ZBS16,Guardado2018,BSM23}, or in an array of optical tweezers, e.g.~\cite{Semeghini2021,Chen2023, Ahn2023, Steinert2023}. 
In these works, spin-motion coupling (arising when the atom explores the spatially-varying potential) is either considered negligible, or as a small source of decoherence with the external d.o.f. treated as a thermal bath. 
For example, if atoms move randomly during the dynamics, because of a finite thermal energy, the interaction varies and blurs the spin dynamics.
By preparing atoms in a pure motional quantum state, the coupling to motion is coherent and creates spin-motion entanglement~\cite{CTM21}. 

In this coherent regime, the spin-motion coupling originates from the variation of the potential $V(r)$ over the rms (root mean squared) spread $x_{\rm rms}$ of the atom position wavefunction, around a distance $d$~\cite{Lesanovsky2023}. The first-order, linear, spin-motion coupling term is parameterized by $\kappa$:
\begin{equation}
\kappa = -x_{\rm rms} \left. \frac{\partial V}{\partial x} \right \vert_{x=d} = 6 \frac{x_{\rm rms}}{d} V(d), \label{eq:spin-motion}
\end{equation} 
where we assumed a repulsive vdW potential. 
First, we compare the ratio of spin-motion to spin-spin coupling $\kappa/V$, which depends on the choice of optical traps: lattice or tweezers. In both approaches, the quantum fluctuation of position $x_{\rm rms} = \sqrt{\hbar/2m\omega}$ ($m$ the mass of the atom) is slightly tunable through the trapping angular frequency $\omega \sim 2\pi \times 10-100$~kHz giving a spread of a few tens of nanometers. The distance $d$ between atoms is typically 0.5~$\mu$m with lattice and can range from 2 to 10~$\mu$m for tweezers. Consequently, the spin-motion coupling is usually only a small perturbation for tweezers $\kappa/V \ll 0.1$~\cite{CTM21}, while it is comparable to the spin-spin coupling in the lattice platform $\kappa/V \sim 0.5$. We will see clear signatures of this large perturbation on the spin dynamics in the first part of this work. 

Secondly, we discuss the relevance of motion through the ratio $\kappa/\omega$, which can vary over many orders of magnitude depending on the platform. For molecules, interacting through a dipole-dipole potential $V$ on the kHz-scale or less~\cite{YMG13,HLC22,BYA22}, the spin-motion coupling is negligible $\kappa / \omega < 0.01$, except if working with delocalized, overlapping, wavefunctions~\cite{LMM23}. For Rydberg atoms excited with cw-lasers, forcing the Rydberg blockade limits the interaction strength $V$ to the MHz-scale which nevertheless allows to enter the perturbative regime $\kappa / \omega \sim 0.1 - 0.5$ and already opens up exciting prospects~\cite{Lesanovsky2020,Lesanovsky2020b,Lesanovsky2023,Mehaignerie2023}. 
By using picosecond pulsed lasers, our ultrafast approach allows to always overcome Rydberg blockade~\cite{MZK20} and prepare Rydberg atoms with interaction strength at the GHz-scale~\cite{TSG16,CTM21,BSM23}. Here, the spin-motion coupling becomes overly dominant with $\kappa / \omega \sim 10 - 1000$, such that motional dynamics, due to the kinetic energy of atoms \cite{ARW07,ARG07,AER08}, can be completely neglected on the timescale of spin-spin and spin-motion entanglement. In the final part of this work, we will propose the ultrafast stroboscopic method to effectively tune $\kappa / \omega$.

\begin{figure}
\centering
\includegraphics[width=.48\textwidth]{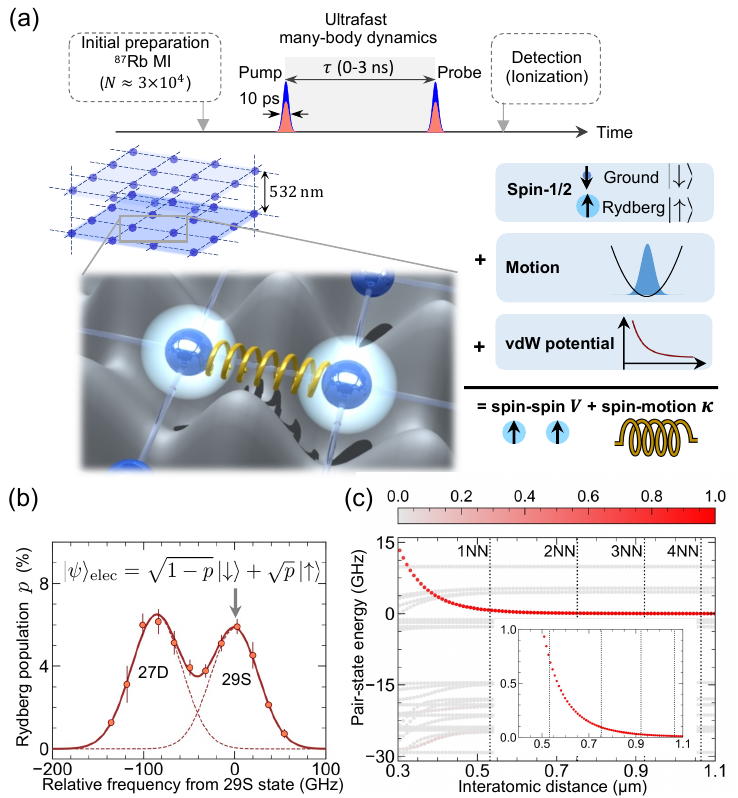}
\caption{
(a) Schematic of the experiment. The atoms prepared in a unity-filling 3D atomic Mott-insulator are coherently excited to the $29S$ Rydberg state using a pump pulse. After the pump excitation, the system undergoes many-body dynamics driven by spin-spin and spin-motion couplings until the probe pulse is applied. 
(b) Rydberg state population after pump excitation. The solid curve shows a fit by a double Gaussian function, while the dashed curves represents the individual contribution from the state 27D and 29S. At resonance, the population in 27D state is only $\sim0.2\%$.
(c) The $29S-29S$ pair state energy as a function of inter-atomic distance. The inset zooms in the $0-1$~GHz energy range.
}
\label{Fig1}
\end{figure}

\textit{Experimental platform} The schematic of our experimental system is shown in Fig.~\ref{Fig1}(a). We prepare a three-dimensional (3D) unity-filling Mott-insulator state with $\sim3 \times 10^4$  atoms in the $\ket{\downarrow} = \ket{5S}$ ground state of $^{87}$Rb (electronic and nuclear spin d.o.f. are fully polarized and decoupled from the ultrafast dynamics). The 3D optical lattice, with period $a_{\rm lat} = 532$~nm, has a depth of $20 \, E_R$ for each axis giving rise to an isotropic trapping frequency $\omega = 2\pi \times 18$~kHz in the harmonic oscillator approximation \cite{BSM23}. The spatial wavefunction $\ket{\psi}_{\rm spatial}$ of each atom, prepared in the motional ground-state of each lattice site, have a quantum uncertainty of position $x_{\rm rms} = 57$~nm, and a momentum uncertainty $p_{\rm rms} = \hbar/2x_{\rm rms} = m \times (6.4$~mm/s). 

Following preparation of the ground-state atoms, they are then coherently excited to the $\ket{\uparrow} = \ket{29S}$ Rydberg state using a two-photon (779 and 483 nm) off-resonant excitation with broadband laser pulses ($\sim$10 picoseconds duration) as described in Ref.~\cite{SM} and shown in Fig.~\ref{Fig1}(b). This prepares each atom in a coherent electronic superposition $\ket{\psi}_{\rm elec} = \sqrt{1-p}\ket{\downarrow} + \sqrt{p}\ket{\uparrow}$, with $p$ the probability to be in the Rydberg state, typically $4-6\%$~\cite{TSG16,BSM23}, and where we mapped the ground and Rydberg states to a spin-1/2. 
Two atoms in the $29S$ state experience strong dipole-dipole interaction in the vdW regime. Figure~\ref{Fig1}(c) shows the interaction potential calculated using the pairinteraction software~\cite{Weber2017,SPA17}. It is very well approximated by an isotropic, repulsive, vdW form $V(r) = C_{6}/r^{6}$, where the calculated coefficient $C_6^{\rm th}$ is $2\pi \times 16 \, {\rm MHz} \, \mu{\rm m}^6$. The mixing with the dominant interaction channel (the pair-state $28P-29P$) remains negligible thanks to its large energy separation of 20 GHz. 
Choosing a Rydberg $S$-state, rather than $D$-state as in previous works~\cite{CTM21,BSM23}, was motivated by obtaining this clean isotropic potential, despite the increased experimental challenge caused by the smaller excitation strength of $S$-state and in spectrally resolving the $S$ and $D$ states when using picosecond laser pulses~\cite{SM}. 

\textit{The model Hamiltonian} Here, we discuss the model Hamiltonian, including the motional d.o.f.. Following excitation, each atom $j$ is initially in a product state of spatial and internal d.o.f. $\ket{\psi_j} = \ket{\psi}_{\rm spatial} \otimes \ket{\psi}_{\rm elec.}$. We then consider the evolution of this system in the nanosecond timescale relevant for spin-spin and spin-motion entanglement. For such short duration, the motion of atoms can be completely ignored: the position probability distribution do not have time to evolve either from the absence of confining potential for the Rydberg state or from the vdW repulsion. The ultrafast dynamics is then driven only by: 

\begin{align}
\frac{\hat{H}}{\hbar} &= \sum_{j<k}V(\hat{\bm{r}}_{jk}) \otimes \hat{n}_j \hat{n}_k \nonumber \\
&\approx \sum_{j<k} \left( V_{jk} + \kappa_{jk} \frac{\hat{\bm{r}}_{jk}  - \bar{\bm{r}}_{jk}}{x_{\rm rms}}\cdot\bm{e}_{jk} + ...\right) \otimes \hat{n}_j \hat{n}_k .
\label{Hamiltonian}
\end{align}
Here, $\hat{\bm{r}}_{jk} = \hat{\bm{r}}_{j} - \hat{\bm{r}}_{k}$ is the quantum operator of the relative position of atoms $j$ and $k$, $\bar{\bm{r}}_{jk}$ its expectation value, $\bm{e}_{jk}$ a unit vector along site $j$ and $k$, $V_{jk}$ and $\kappa_{jk}$ the couplings evaluated at distance $\bar{r}_{jk}$, and $\hat{n}_j=\ket{\uparrow}_j\bra{\uparrow}$ is the projection operator on the Rydberg state for the $j$-th atom. Applying this Hamiltonian to the initial product state creates entanglement within the spin sector, but also, and this is the key point of the first part of this work, between the spin and motional sectors of the Hilbert space.

\begin{figure}
\centering
\includegraphics[width=.48\textwidth]{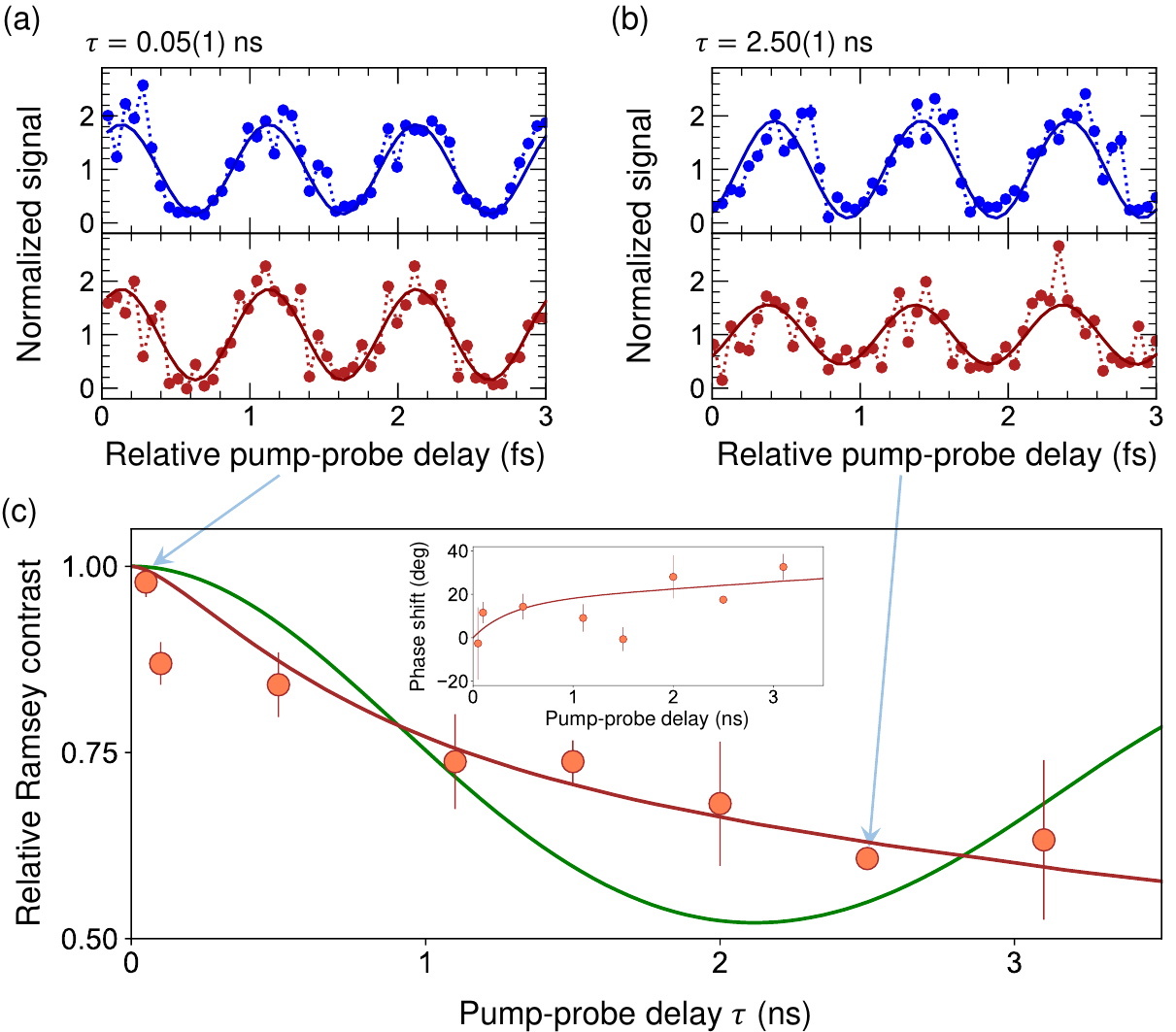}
\caption{Time-domain Ramsey interferograms for atoms prepared in a Mott-insulator, strongly-interacting, state (red) and a low-density, non-interacting, reference atomic sample (blue) for (a)~$\tau=0.05$~ns and (b)~$\tau=2.5$~ns. The vertical axis of each interferogram is normalized by the mean value of data. (c) Measured relative Ramsey contrasts (ratio of the contrasts of Mott-insulator and reference sample) and phase shifts (phase difference of Mott-insulator and reference sample) are shown by red circles (error bars are standard error of the mean). The observations are compared with a fitted numerical solution that takes into account spin-motion coupling (red curve) or ignore it (green curve).} 
\label{Fig2}
\end{figure}

\textit{Results} We now present experimental results obtained by time-domain Ramsey interferometry~\cite{TSG16,BSM23,CTM21} with $p\sim 4.8\%$, to probe the many-body entangled state generated by the above Hamiltonian. 
In short, a first pump pulse initiates the many-body dynamics which is read-out by a second probe pulse after a variable delay $\tau = 0 - 3$~ns. 
This second pulse gives rise to a Ramsey interference whose contrast is a probe to the single-atom coherence in the spin sector, i.e., between the ground and Rydberg state. 
Spin-spin and spin-motion coupling generates entanglement entropy~\cite{BSM23}, which reduces the single-atom coherence and thus the Ramsey contrast. Ramsey interferograms are obtained by measuring the Rydberg population \textit{p} after the probe pulse, as a function of relative pump-probe delay, by detecting the field-ionized Rybderg atoms using a microchannel plate \cite{SM}.
Typical interferograms are shown in Fig.~\ref{Fig2}(a,b). In absence of interaction (blue curve, obtained for a low-density atomic sample), the highly contrasted interference indicates a constant pure state. 
For atoms prepared as a Mott-insulator (red curve), the decreasing contrast signals a reduced purity in the spin sector, which is shown in Fig.~\ref{Fig2}(c) as a function of the delay $\tau$. Additionally, we also extract a phase shift of the Ramsey oscillations with the reference non-interacting sample. 

\textit{Numerical solution} To calculate the Ramsey contrast and phase shift from the action of the Hamiltonian of Eq.~(\ref{Hamiltonian}), we extend previous results~\cite{TSG16,SPT16} to include the spatial wavefunction of each atom, which requires to calculate terms such as the two-body spatial overlap:
\begin{align}
O_{jk}(t) &= \langle \psi_j; \psi_k | \exp(-iV(\hat{\bm{r}})t) | \psi_j; \psi_k \rangle \nonumber \\
& = C \int d\bm{r} \,  \exp\left(- \frac{|\bm{r} - \bar{\bm{r}}_{jk}|^2}{x_{\rm rms}^2} -i \frac{C_6}{|\bm{r}|^6}t \right),
\label{overlap}
\end{align}
where $C$ is a normalization constant.
The second line is obtained after reformulating the two-body wavefunctions $\ket{\psi_j;\psi_k}$ into two independent one-body system: a trivial one for the center-of-mass, unaffected by the interaction, and the interesting one for the relative coordinate $\bm{r}_{jk}$ with reduced mass $m/2$. For a two-atom system, the Ramsey contrast and phase are directly related to the amplitude and phase of the complex-valued overlap $O$. For the many-body dynamics considered here, the analytical expression relating them is given in Ref. \cite{SM}, which also include details on neglecting three-body (and higher) overlap terms.

The calculation results are then fitted to the relative Ramsey contrast data with a single free parameter: the coefficient $C_6$. The fitted curve, see Fig.~\ref{Fig2}(c), agrees well with the experimental data for a coefficient $C_6^{\rm exp} = 2\pi \times  5.5 \, {\rm MHz} \, \mu{\rm m}^6$. With this value, the positive trend (related to the sign of $C_6$) and magnitude of the phase shift are also well captured. 
The fitted $C_6^{\rm exp}$ coefficient is $3$ times smaller than obtained from ab-initio calculation of the vdW potential, which calls for further investigation of the accuracy of the vdW potential calculation in the short, sub-micron distance regime. This could be done using a tweezers platform where a simpler system of only two atoms can be prepared~\cite{CTM21}, potentially down to the short sub-micrometer distance by throwing atoms with moving tweezers~\cite{HBP23}.

To emphasize the importance of the spin-motion coupling in this experiment, we also show calculation for a pure spin-spin model where we ignore the spatial extent of the wavefunctions \cite{TSG16,BSM23}. As shown in the green curve of Fig.~\ref{Fig2}(c), the Ramsey contrast would have displayed an oscillation (see \textit{Discussion}) which is clearly absent in the experimental data.
We can thus conclude that capturing spin-motion entanglement is essential to account for the observed many-body dynamics.

\begin{figure}
\centering
\includegraphics[width=.48\textwidth]{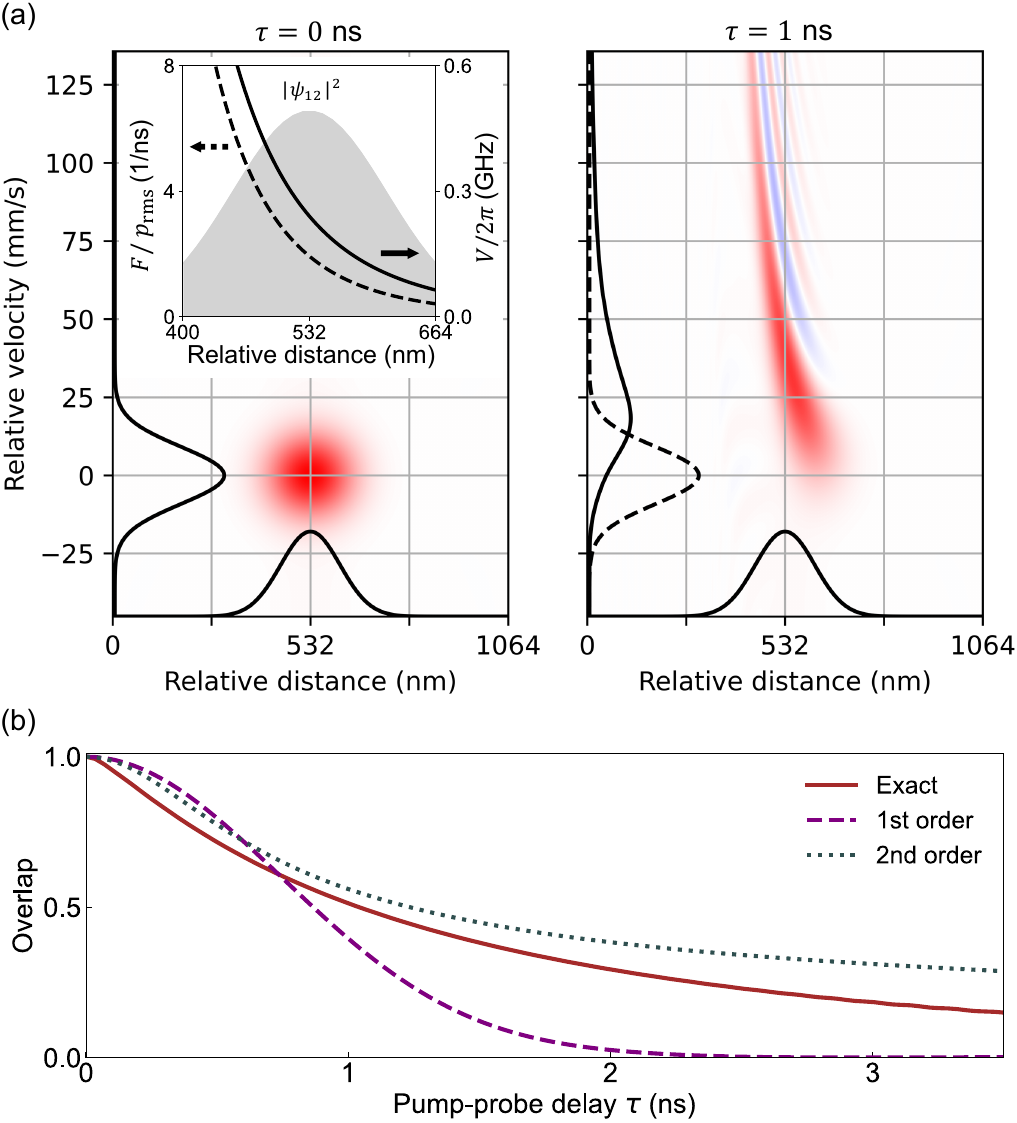}
\caption{(a) Phase-space (Wigner) representation of the relative wavefunction $\psi_{12}$ at time $\tau=$ 0 and 1~ns.  The $\psi_{12}$ at $\tau=0$~ns is the relative wavefunction of the two atoms in the motional ground state of the lattice sites. The red (blue) color represents positive (negative) value of the Wigner distribution. The marginal position and velocity distribution are shown as black lines. The momentum displacement and squeezing are clearly visible. Inset: zoom on the probability distribution $|\psi_{12}|^2$, showing the spatial variation of the vdW potential $V$ (solid), and the resulting force $F$ (dashed). (b) Overlap $|O(t)|$ as a function of the delay $\tau$. Solid curve: exact calculation of Eq.~(\ref{overlap}). The dashed (dotted) lines are obtained by expanding the vdW potential to first-order (second-order).}
\label{Fig3}
\end{figure}

\textit{Discussion} 
We now present a hierarchy of approximations to identify the relevant terms in Eqs.~(\ref{Hamiltonian},\ref{overlap}) that create spin-motion entanglement. We consider two atoms at nearest-neighbour (NN) distance $a_{\rm lat}$, where the variation of potential over the wavefunction describing their relative distance $\psi_{12}(\bm{r})$ is largest. 
We then restrict the problem to 1D, along the inter-atomic axis, by neglecting the wavefunction spread in the other two directions as it gives a small $\frac{1}{2}(x_{\rm rms}/a_{\rm lat})^2 \simeq 0.5~\%$ increase in NN distance, 20 times smaller than the effect along the inter-atomic axis. This allows a phase-space representation of the 1D wavefunction $\psi_{12}(x)$, as shown in Fig.~\ref{Fig3}(a), which is convenient to depict the relative motional states of two atoms with the Rydberg interaction \cite{Mehaignerie2023}.

The $1/r^6$-potential then applies a strong force on the wavefunction which can be decomposed with a series expansion of the potential around the mean interatomic distance $a_{\rm lat}$. The zeroth-order term $V = C_6^{\rm exp}/a_{\rm lat}^6$ gives rise to spin-spin entanglement reaching its maximal value at time $\tau = \pi/V = 2.1$~ns, and corresponding to a minimum in the Ramsey contrast of the green curve of Fig.~\ref{Fig2}(c). For longer time, the two effective spins would de-entangle and the Ramsey visibility restore~\cite{Ahn2020,CTM21}. The first-order linear term, explicitly written in Eq.~(\ref{Hamiltonian}), gives a uniform force on the wavefunction $F = 6 \hbar C_6/a_{\rm lat}^7 = \hbar \kappa / x_{\rm rms} \simeq \frac{m}{2} (2.5 \times 10^7 \, {\rm m \, s}^{-2})$. The momentum kick $\Delta p$ from this acceleration becomes comparable to the relative momentum rms spread after $\tau = p_{\rm rms}/\sqrt{2}F = 0.3$~ns. As the state-dependent force is applied only on part of the spin sector ($\ket{\uparrow \uparrow}$), it creates spin-motion entanglement that is captured by the reduced overlap $|O|$
between the displaced and initial momentum wavefunction seen in Fig.~\ref{Fig3}(b). It explains why the Ramsey contrast drops initially faster than expected from a pure spin model, see Fig.~\ref{Fig2}(c), as well as why it does not restore beyond $\tau = 2.1$~ns as the pure spin model predicts.

For a good qualitative description of the dynamics, it is necessary to go beyond the first-order term to capture the wide variation of the mechanical force over the wavefunction. As seen in Fig.~\ref{Fig3}(b), a second-order expansion brings the calculated overlap much closer to the exact result from Eq.~(\ref{overlap}). Qualitatively, these second-order terms $\hat{r}_{jk}^2 = (\hat{x}_j-\hat{x}_k)^2$ have two interesting effects on the wavefunction. First, they squeeze each atom wavefunction through the terms $\hat{x}_j^2$ and $\hat{x}_k^2$: the atoms feel a stronger force at shorter distance from the other one, which will compress the wavefunction. And secondly, they entangle the two atoms wavefunctions through the cross term $\hat{x}_j \hat{x}_k$. The relative wavefunction $\psi_{12}$ cannot anymore be decomposed into a product state of two single-atom wavefunctions. Such entanglement between the motion of two atoms is not captured at lower order. The third-order terms are required to explain the negative value taken by the Wigner distribution. 

\textit{Outlook} The strong spin-motion coupling observed here precludes the realization of a pure spin model in our experimental regime. However, instead of performing quantum simulation in the spin sector, we could rather work fully in the motion sector of the Hilbert space. This would be realized by completely transferring ground-state atoms to Rydberg orbits, a step that can be done with high-fidelity in the microsecond timescale~\cite{Levine2018} (but only for weakly interacting atoms), and for which progress have been reported by our group for picosecond-scale excitation~\cite{CTM21}. We could then prepare a unit-filling Mott-insulator state of \textit{Rydberg} atoms which would be submitted to strong internal vdW force~\cite{Morsch2016}. Interestingly, the forces from two opposite directions of a given atom cancel in first-order and the second-order squeezing and entangling terms would dominate the dynamics. This would lead to non-trivial distortion of the spatial wavefunctions observable by time-of-flight imaging, a technique also available on the tweezers platform~\cite{Bergschneider2019,Brown2022}. 

\begin{figure}
\centering
\includegraphics[width=.48\textwidth]{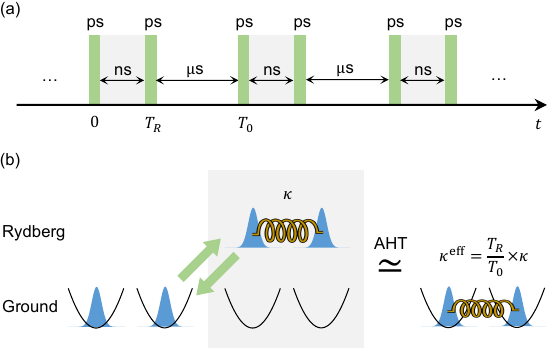}
\caption{Hamiltonian engineering by \textit{ultrafast stroboscopic Rydberg excitation}. (a) Pulse sequence, see text. (b) The stroboscopic sequence gives rise to an average Hamiltonian with tunable interaction strength competing with the trapping potential.}
\label{Fig4}
\end{figure}

In this work, we neglected the effect of kinetic energy due to the large separation of timescales between the Rydberg interaction (nanoseconds) and the motion of atoms (microseconds). We now propose to bring these two scales together to investigate a larger class of Hamiltonians with \textit{ultrafast stroboscopic Rydberg excitation}. As schematically drawn in Fig.~\ref{Fig4}, ground-state atoms are transferred in a picosecond-timescale to Rydberg states to experience for a brief time $T_R$ the strong force demonstrated in this work. This gives a momentum kick that can be widely tuned, by $T_R$ and the choice of Rydberg state, with respect to the trap depth (we should not kick atoms out of their trapping sites). They are then brought back to the ground-state to now experience the kinetic energy and the trapping potential on a microsecond-timescale $T_0$. This step is repeated with a high enough frequency to apply Average Hamiltonian Theory (AHT)~\cite{AHT1968,Dalibard2014,Eckardt2015,Choi2020}, and a controlled duty cycle to vary the effective, reduced, coupling strength $\kappa_{\rm eff} = T_R/T_0 \times \kappa$ relatively to the trapping frequency $\omega$. Optionally, a spin-1/2 can be encoded in the ground-state manifold, and a spin-dependent force obtained by spin-selective ultrafast excitation. We note that this requires to combine ultrafast excitation with resolving the 6.8~GHz = 1/(150~ps) hyperfine splitting, which is one of our ongoing developments~\cite{subns_5P}.

This ultrafast Floquet engineering approach can be seen as complementary to Rydberg dressing~\cite{JHK16,ZBS16,SchleierSmith2020,GSB21,Steinert2023,Hines2023}, where a trapped ground-state atom is instead continuously and weakly dressed by a small fraction of Rydberg character. Compared to other proposals for spin-motion coupling using long-lived circular Rydberg states~\cite{Mehaignerie2023}, or Rydberg facilitation (anti-blockade)~\cite{Lesanovsky2020,Lesanovsky2020b,Lesanovsky2023}, here we note that the stroboscopic approach have the practical advantage to not require magic-trapping of the Rydberg state. Finally, we emphasize that ultrafast Rydberg excitation with pulsed lasers (delivering up to 100 GHz of ground-Rydberg Rabi frequency) unlocks the full GHz-strength of interaction between Rydberg atoms, otherwise curbed by the limited MHz-scale Rabi frequency achievable with cw-lasers. 

In conclusion, we have considered the force experienced by Rydberg atoms, mapped it into a spin-motion coupling term, and observed a clear signature: a strong perturbation to the spin dynamics. We proposed a quantum control technique, ultrafast Floquet engineering, to tune the relative strength of this force compared to the trapping potential of optical lattice or tweezers, opening novel regimes of quantum simulation with Rydberg atoms. Among the new avenues, we envision the creation of exotic motional states such as a Rydberg crystal: an atomic array with each atom stabilized in free-space (i.e., in the absence of a confining lattice potential) by long-range isotropic vdW repulsion between Rydberg atoms, a state reminiscent of electronic Wigner crystals~\cite{Wigner1934}. 

The authors acknowledge Y. Okano and H. Chiba for the technical support. We thank E. Braun, H. Tamura, A. Trautmann, S. Weber, C. A. Weidner for fruitful discussions; and Y. Zhang for the helpful discussions regarding the extension of delay line. This work was supported by MEXT Quantum Leap Flagship Program (MEXT Q-LEAP) JPMXS0118069021, JSPS Grant-in-Aid for Specially Promoted Research Grant No. 16H06289 and JST Moonshot R\&D Program Grant Number JPMJMS2269. S.S. acknowledges support from JSPS KAKENHI Grant No. JP21H01021. M.K. acknowledges supports from JSPS KAKENHI Grants No. JP20K14389 and No. JP22H05268.

Note added. Recently we became aware of related work on spin-motion entanglement that demonstrates quantum information processing using motional d.o.f. in tweezers \cite{SSR23}.


%


\newpage

\renewcommand{\theequation}{S\arabic{equation}}
\renewcommand{\thefigure}{S\arabic{figure}}
\renewcommand{\bibnumfmt}[1]{[S#1]}
\renewcommand{\citenumfont}[1]{S#1}

\section{Supplemental material}

\subsection{Rydberg excitation}
The unity-filling atomic Mott-insulator is prepared in the $5S_{1/2}, \ket{F=2, m_F=-2}$ hyperfine ground state of $^{87}$Rb \cite{BSM23_1}. We turn off the trapping and optical lattice beams $\sim$~2~$\mu$s before the Rydberg excitation to avoid multiphoton ionization. We then use two-photon excitation, with picosecond infrared (IR) and blue laser pulses, to excite the ground-state atoms to the $\ket{29S_{1/2}, m_F=-2}$ Rydberg state. The pulsed laser system for the excitation is as described in ref. \cite{BSM23_1}.

\begin{figure}
\centering
\includegraphics[width=.4\textwidth]{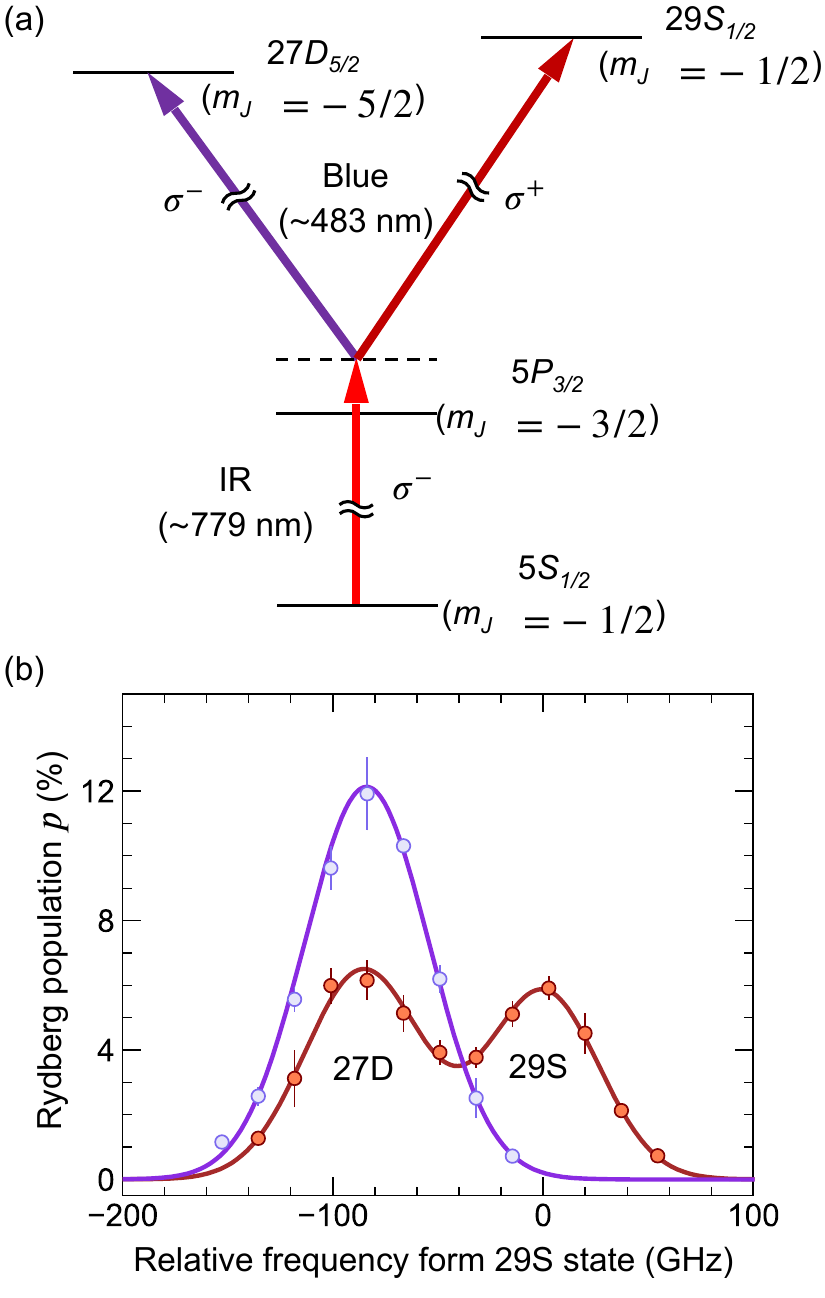}
\caption{(a) Scheme for two-photon excitation of 29S and 27D Rydberg states. (b) Example of excitation bandwidth measurement using 27D state (purple data and Gaussian fit). The red data and fit are the same as Fig~1(b).}
\label{Excitation}
\end{figure}

In detail, we prepare the 29S Rydberg state by using $\sigma^-$ and $\sigma^+$ polarized IR and blue pulses as shown in Fig.~\ref{Excitation}(a). In order to resolve the 29S state from the nearby 27D state, which is only 80 GHz lower in energy, we reduce the excitation bandwidth from previous works by roughly half. At the 29S state resonance (where we perform experiments), the population in the nearby 27D state is found to be only 4~\% of the total population which is negligibly small. The laser pulses have an energy of 50~nJ (IR) and 560~nJ (blue), and  a $1/e^2$-diameter of $230$~$\mu$m (IR) and $50$~$\mu$m (blue). 

In order to estimate the excitation bandwidth, we excite only the $\ket {27D_{5/2}, m_F =  -4}$ state by using $\sigma^-$-polarization for both laser pulses (Fig.~\ref{Excitation}(a)). A fit to the purple curve in Fig.~\ref{Excitation}(b) gives a 72(5)~GHz (FWHM) bandwidth. In this measurement, the blue pulse energy was reduced to 300~nJ.

\subsection{Rydberg state detection}
At the end of the experiment, the Rydberg atoms are ionized by a strong electric field, detected by a micro-channel plate (MCP), and counted after going through a pre-amplifier and a time-gated integrator. This detection setup is as described in ref. \cite{BSM23_1}. 
For the field ionization, we applied $+$2.5 kV pulses to six electrodes (red electrodes in ref. \cite{MZK20_1}) and $-$3 kV pulses to two electrodes (blue electrodes in ref. \cite{MZK20_1}). 

\begin{figure}
\centering
\includegraphics[width=.45\textwidth]{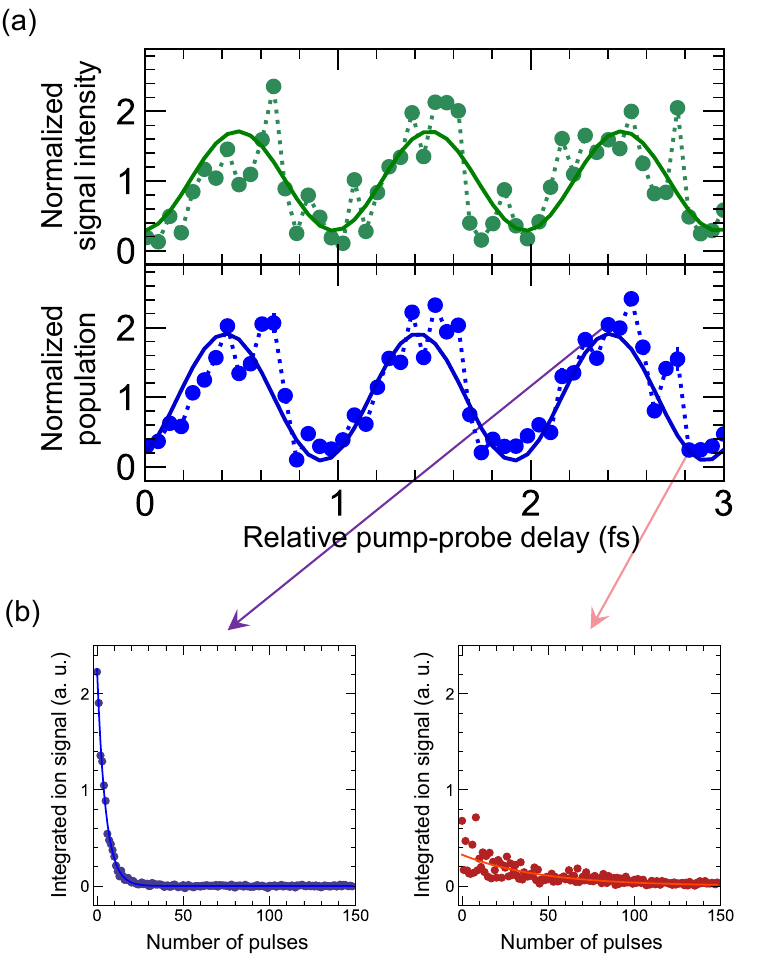}
\caption{The Ramsey measurement for reference sample at $\tau \sim$~2.5~ns. (a) The green points show the measured Rydberg population obtained after one pump-probe pulse, while the blue points are obtained by fitting an exponential decrease, as desribed in the text. The error bars in blue points show standard deviation. The solid curves show the sinusoidal fit to the data. (b) Examples of Rydberg population estimation at the maximum and miminum of the Ramsey fringe.}
\label{LDpop}
\end{figure}

\subsection{Ramsey measurements}

The Ramsey measurements are performed by producing a pair of pump and probe pulses with an optical delay-line interferometer \cite{BSM23_1}. The pump-probe delay $\tau$ was tuned by a mechanical stage, whereas the fine delay was controlled with attosecond (as) precision by using a piezoelectric transducer to observe the 1~femtosecond-period Ramsey fringe. We scanned the relative delay in steps of $\sim 60$~as over a range of $\sim 3$ femtoseconds, see Fig.~\ref{LDpop}(a).

Experiments are realized on both a Mott-insulator sample, where atoms strongly interact with their neighbours, as well as on a low-density reference sample where interaction can be neglected. This procedure is described in ref. \cite{BSM23_1}, and allows to extract a phase shift of the Ramsey interferograms between the interacting and reference sample. For the Mott-insulator measurement, the Rydberg population is obtained after sending a single pump-probe pulse for each experimental realization. For the reference measurement, we repeat the pump-probe sequence 150 times, every 1 millisecond, on each low-density sample. This allows to increase statistics and reduce the influence of shot-to-shot uncertainties in pulse energy, pulse pointing and atom number fluctuations. The ion signals are measured after each pair of pump and probe pulses and we record its decrease caused by a depletion of the sample from the finite Rydberg population. This population is extracted by fitting an exponential decay \cite{MZK20_1}. By implementing this scheme, there is 30~\% reduction in statistical uncertainties for the estimation of contrast and phase of reference sample Ramsey interferograms, as shown in Fig.~\ref{LDpop}.

The Ramsey signals are measured alternately for the reference sample and the Mott-insulator state. Before and after each measurement, we check the number of atoms $N$ and Rydberg state population $p$. We exclude the data with deviations of average values by more than 15~\% from the set values: $N^{\rm{set}}\sim 30000$ atoms and $p^{\rm{set}}\sim 4.8\%$, and 15~\% change, as compared to average values, in values recorded after measurement.

\subsection{Many-body dynamics for spin-motion coupled system}
The Ramsey signal $P_j$ for an atom $j$ is given by the following expression \cite{SPT16_1}:
\begin{equation}
    P_{j}(\tau) = 2p (1-p) \mathrm{Re}[1+C_j(\tau) e^{(iE_{r}\tau/\hbar + \phi_0)}]
\end{equation}
Here, $p$ is the population in the Rydberg state, $E_r$ is the energy difference between the ground and Rydberg state, $\tau$ is the pump-probe delay, $\phi_0$ is a phase arising from the AC-Stark shifts during the pulse excitation, and $C_j(\tau)$ is interaction-induced modulation of the Ramsey fringe that reflects the coherences established in the system during the many-body dynamics. The experimentally observed many-particle signal is given by averaging over contributions from all the atoms $\bar{P}(\tau) = (1/N)\sum_{j = 1}^{N} P_{j}(\tau)$. 

The term $C_j(\tau)$ contains the full signature of the interactions, and Ramsey contrast and the phase are related to its absolute value and angle, respectively. For a pure spin-spin model ($S\mbox{-}S$) and just two atoms $j$ and $k$, this term reads: 
\begin{equation}
C^{S\mbox{-}S}_{j,k}(\tau) =  (1-p) + p e^{i V_{jk} \tau},
\end{equation}
where $V_{jk}$ is the van der Waals potential between atom $j$ and $k$. This complex-valued term has minimum amplitude for $V_{jk}\tau = \pi$, corresponding to maximal spin-spin entanglement. In the special case $p = 0.5$, the Ramsey constrast would vanish. 
For a many-body system, interaction of atom $j$ with all possible other atoms $k$ has to be included \cite{SPT16_1}: 
\begin{equation}
C^{S\mbox{-}S}_j(\tau) = \prod_{k \neq j} \left[ (1-p) + p e^{i V_{jk}\tau} \right].
\end{equation}

We now include the external degrees of freedom and spin-motion coupling ($S\mbox{-}M$). We first focus on a case of only two atoms $j$ and $k$ and obtain:
\begin{align}
C^{S\mbox{-}M}_{j,k}(\tau) &= (1-p) + p \langle \psi_j; \psi_k | e^{i V(\hat{\bold{r}})\tau} | \psi_j; \psi_k \rangle, \\
&= (1-p) + p O_{jk}(\tau)
\end{align}
where $|\psi_j\rangle$ describes the spatial wavefunction of atom $j$, and $O_{jk}(\tau)$ is the overlap term introduced in the main text. For atoms localized to an infinitesimal region, the overlap reduces to the previous case $O_{jk}(\tau) = e^{iV_{jk} \tau}$. 

Extending the calculation to the many-body spin-motion-coupled system is more subtle that for the spin-spin model. Indeed, a strict derivation requires the calculation of terms of higher-order, such as $O_{jkl} = \langle \psi_j; \psi_k; \psi_l | e^{i V(\hat{\bold{r}})\tau} | \psi_j; \psi_k; \psi_l \rangle$, corresponding to 3 atoms $j$, $k$ and $l$ in the Rydberg state. Such higher-order terms do not decompose simply into product of two-body terms, as for the pure spin model. The physical picture being that the momentum kicks on atom $j$ from two other atoms $k$ and $l$ can compensate each other. However, to simplify the calculations, we perform the approximation that higher-order overlaps decomposes into products of two-body overlaps and write: 
\begin{equation}
C^{S\mbox{-}M}_j(t) \simeq \prod_{k \neq j} [(1-p)+ p O_{jk}(t)].
\end{equation}
We justify this approximation by pointing out that, in our regime of low Rydberg state population ($p\sim4.8\%$), the dominant error term (3-body overlaps) contributes to a negligibly small fraction $p$ of the two-body terms. Finally, the calculation over other atoms $k$ is performed only up to the fourth nearest-neighbor in the 3D lattice (distance $2 a_{\rm lat}$), where the interaction has already dropped by a factor $2^6 = 64$.

\end{document}